\documentclass[twocolumn,showpacs,twoside,10pt,pra]{revtex4}
\usepackage{multirow,amsthm,amsmath,amssymb,mathrsfs}
\usepackage{epsfig,epsf,graphicx,subfigure}
\usepackage[colorlinks=true,linkcolor=blue,anchorcolor=blue,citecolor=blue,urlcolor=blue]{hyperref}
\usepackage{appendix}
\newcommand{\bra}[1]{\langle #1|}
\newcommand{\ket}[1]{|#1\rangle}
\usepackage[usenames]{color}
\definecolor{Red}{rgb}{1,0,0}
\definecolor{Blue}{rgb}{0,0,1}

\begin{document}

\title{Experimental observation of information flow in the anti-$\mathcal{PT}$-symmetric system}

\author{Jingwei Wen$^{1}$}
\author{Guoqing Qin$^{1}$}
\author{Chao Zheng$^{5}$}
\author{Shijie Wei$^{1,4}$}
\author{Xiangyu Kong$^{1}$}
\author{Tao Xin$^{6}$}
\email{xint@sustc.edu.cn}
\author{Guilu Long $^{1,2,3}$}
\email{gllong@tsinghua.edu.cn}

\affiliation{$^{1}$ State Key Laboratory of Low-Dimensional Quantum Physics and Department of Physics, Tsinghua University, Beijing 100084, China}
\affiliation{$^{2}$ Tsinghua National Laboratory for Information Science and Technology,  Beijing 100084, P. R. China.}
\affiliation{$^{3}$ Collaborative Innovation Center of Quantum Matter, Beijing 100084, China}
\affiliation{$^{4}$ Beijing Academy of Quantum Information Sciences, Beijing 100193, China}
\affiliation{$^{5}$ Shenzhen Institute for Quantum Science and Engineering, Southern University of Science and Technology, Shenzhen 518055, China}
\affiliation{$^{6}$ Center for Quantum Computing, Peng Cheng Laboratory, Shenzhen 518055, China}


\begin{abstract}
The recently theoretical and experimental researches related to $\mathcal{PT}$-symmetric system have attracted unprecedented attention because of various novel features and potentials in extending canonical quantum mechanics. However, as the counterpart of $\mathcal{PT}$-symmetry, there are only a few researches on the anti-$\mathcal{PT}$-symmetric quantum system because of the challenges in the quantum simulation of anti-$\mathcal{PT}$-symmetry in experiment. Here, we propose an algorithm for simulating the generalized anti-$\mathcal{PT}$-symmetric system with quantum circuit. Utilizing the protocols, an oscillation of information flow is observed for the first time in our Nuclear Magnetic Resonance quantum simulator. We show that the information will recover from the environment completely when the anti-$\mathcal{PT}$-symmetry is broken, whereas no information can be retrieved in the symmetry-unbroken phase. Our work opens the gate for the practical quantum simulation and experimental investigation of universal anti-$\mathcal{PT}$-symmetric system in the quantum computer.
\end{abstract}

\pacs{03.67.Ac, 03.67.Lx, 76.60.-k,03.65.wj}
\maketitle

\section{Introduction}
The limitation on the Hermiticity of Hamiltonian can ensures the reality in the spectrum of energy eigenvalues and the unitarity of the resulting time evolution. However, a new class of non-hermitian Hamiltonian has attracted extensive attention and research because of the discovery by Bender and Boettcher in 1998 \cite{Bender}. It was found that Hamiltonians satisfying parity $\mathcal{P}$ (spatial reflection) and $\mathcal{T}$ (time reversal) symmetry instead of hermiticity can still have real energy spectra and orthogonal eigenstates in the symmetry-unbroken phase, in which the eigenfunction of system Hamiltonian is at the same time an eigenfunction of the joint $\mathcal{PT}$ operator \cite{nonlinear,add_pt}. When the Hamiltonian parameters cross the exceptional point, $\mathcal{PT}$-symmetry will be broken and lead to a symmetry-breaking transition \cite{EP1,cir_EP,pt_the1}. This work has inspired numerous theoretical and experimental studies \cite{exp_pt0,exp_pt1,exp_pt2,exp_pt3,exp_pt4,exp_pt5,flow} of the non-hermitian systems, including demonstrating novel properties of quantum systems \cite{violation,entangle1} and extending fundamental quantum mechanics \cite{reconstruct,reconstruct2}. 

However, there are limited investigations on another important counterpart anti-$\mathcal{PT}$-symmetry, which means the system Hamiltonian is anti-commutative with the joint $\mathcal{PT}$ operator $\{H,\mathcal{PT}\}=0$. Some relevant experimental demonstrations have been realized in atoms \cite{antipt_atoms,antipt_atoms2,antipt_atoms3}, optical \cite{antipt_optica0,antipt_optica,antipt_optica2,antipt_optica3,antipt_optica4,antipt_optica5}, electrical circuit resonators \cite{antipt_resonc} and diffusive systems \cite{antipt_diff}. Quantum processes and quantum properties such as symmetry breaking transition, observation of exceptional point, refractionless propagation, and simulation of anti-$\mathcal{PT}$-symmetric Lorentz dynamics have been presented in these experiments \cite{antipt_atoms,antipt_optica,antipt_optica2,antipt_optica5,antipt_resonc}, whereas the novel characteristics of entanglement \cite{entangle1,entangle2,entangle3} and information flow \cite{flow,flowpra,flowpra2} in the anti-$\mathcal{PT}$-symmetric system, which would present various phenomena different from Hermitian quantum mechanics and reveal the relationship between system and environment, have not been fully and thoroughly investigated in the experiment. 

In this work, we propose an algorithm for the simulation of generalized anti-$\mathcal{PT}$-symmetric evolution with quantum circuit model and report the first experimental observation of information flow oscillation in anti-$\mathcal{PT}$-symmetric system on Nuclear Magnetic Resonance quantum computing platform. The simulation scheme is based on decomposing the non-Hermitian Hamiltonian evolution into a sum of unitary operators and realizing the simulation in an enlarged Hilbert space with ancillary qubits \cite{enlarge,dqc1,lcu}. We experimentally show that the information flow oscillates back and forth between the environments and system in broken phase and the phenomenon of information backflow occurs, which indicates information flows from the environment back to the system and this does not happen in traditional Hermitian quantum mechanics \cite{flow,monotonicity1,monotonicity2}. The oscillation period and amplitude increase as the system parameters approach the exceptional point. When passing through the critical point, the information backflow no longer occurs, and can only be attenuated exponentially from the system and leakage information into the environment. The monotone correspondence relation can provide a way to measure the degree of hermiticity for a quantum system. The phase-breaking and information flow transition at exceptional point observed in our experiment also mean the transition from non-Markovian process to Markovian process \cite{flowpra2,Markovian1,Markovian2}. 

\section{Construction of Simulation Algorithm} 
We start from a more generalized form for a single-qubit anti-$\mathcal{PT}$-invariant Hamiltonian, which can be expressed as
\begin{equation}
\hat{H}=
\begin{pmatrix}
re^{i\theta}&is\\
i\mu&-re^{-i\theta}\\ 
\end{pmatrix} 
\label{eq1}
\end{equation}
where all the parameters $r$, $\theta$, $s$, $\mu$ are real numbers \cite{reconstruct}. This generalized Hamiltonian will be the negative transpose of the original form under the action of joint operator $\mathcal{PT}$, where operator $\mathcal{P}$ is the Pauli $\sigma_{x}$ matrix and $\mathcal{T}$ corresponds to complex conjugation. When the condition $s=\mu$ is satisfied as well, it can be reduced to the anti-commutation relation 
\begin{equation}
\begin{split}
\mathcal{(PT)}\hat{H}\mathcal{(PT)}^{-1}=-\hat{H}^{\textup{T}}=-\hat{H}
\end{split}
\end{equation}
where notation $A^\textup{T}$ means the transpose of matrix $A$. The eigenvalues of Hamiltonian $\hat{H}$ are $\varepsilon_{\pm}= ir\sin \theta \pm \sqrt{r^{2}\cos^{2}\theta-\mu s}$ and the system is termed in the regime of unbroken anti-$\mathcal{PT}$-symmetric phase when $r^{2}\cos^{2}\theta-\mu s<0$. For convenience, we set the difference between the two eigenvalues as $w=\varepsilon_{+}-\varepsilon_{-}=2\sqrt{r^{2}\cos^{2}\theta-\mu s}$. The dynamic evolution governed by the non-Hermitian Hamiltonian in Eq.(\ref{eq1}) can be described by

\begin{equation}
\hat{\rho}(t)=\frac{e^{-i\hat{H}t}\hat{\rho}(0)e^{i\hat{H}^{\dagger}t}}{tr[e^{-i\hat{H}t}\hat{\rho}(0)e^{i\hat{H}^{\dagger}t}]} 
\label{rho1}
\end{equation}

Here we employ the usual Hilbert-Schmidt inner product instead of a preferentially selected one and consider the effective non-unitary dynamics of an open quantum system \cite{flow,inner}. Suppose that the non-unitary evolution operator $U$ can be decomposed into the form $U=\sum_{i=1}^{d}\alpha_{i}U_{i}$, where $U_{i}$ are unitary oprators and coefficients $\alpha_{i}$ depend on the choice of decomposition but satisfy construction condition $\sum_{i=1}^{d}\vert \alpha_{i} \vert=\alpha$. Then, by constructing operators $\hat{G}=\sum_{i=1}^{d}\sqrt{\frac{\alpha_{i}}{\alpha}}\ket{i}\bra{0}$ working on the ancillary qubits and controlled-operator $\hat{U}=\sum_{i=1}^{d}\ket{i}\bra{i}\otimes U_{i}$, we can realize the simulation of Hamiltonian evolution in the subspace of $\bra{0}\hat{G^{\dagger}}\hat{U}\hat{G}\ket{0}$ \cite{dqc1,dqc2,dqc3,epl}. We construct a general quanutm circuit to simulate the dynamics evolution in Eq.(\ref{rho1}) by enlarging the system with entangled ancillary qubits and encoding the subsystem with the non-Hermitian Hamiltonian with post-selection.

\begin{figure}
          \centering
  	  \includegraphics[scale=0.8]{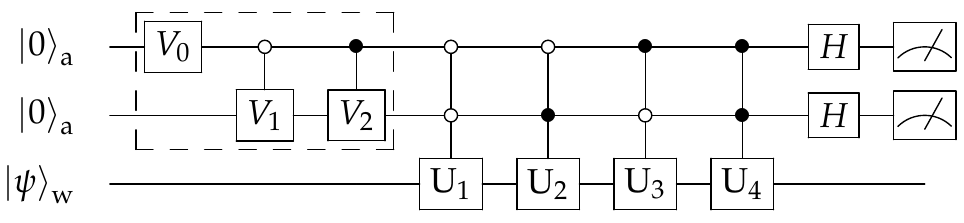}
          \caption{Quantum circuits for the simulation of a generalized anti-$\mathcal{PT}$-symmetric single-qubit system. Single-qubit operators $V_{0}$ and $H$ (Hardmard gate) are operated on the ancillary qubits and two two-qubit operators (0-controlled $V_{1}$ and 1-controlled $V_{2}$) are implemented to the system followed by four three-qubit controlled quantum gates $U_{i} (i \in [1,2,3,4])$. At the end of the circuit, we measure the state vector of the work system in the subspace where the ancillary qubits are $\ket{00}\bra{00}$.}
           \label{fig11:cir3.pdf} 
\end{figure}

Quantum circuit used to simulate the generalized anti-$\mathcal{PT}$-symmetric evolution is shown in Fig. \ref{fig11:cir3.pdf} including two ancillary qubits and one work qubit forming a three-qubit scheme. The initial state is prepared in $\ket{00}_{\textup{a}}\ket{\psi}_{\textup{w}}$ first, then an unitary operator $V_{0}$ and two two-qubit operators (0-controlled $V_{1}$ and 1-controlled $V_{2}$) are implemented on the ancillary qubits just as shown in the dotted box. The operation in the dotted box is equivalent to an operator $V$ and only the first column is definable. The first column of two-qubit operator $V$ without considering the normalization constant is $[V_{11},V_{21},V_{31},V_{41}]$ and

\begin{equation}
\begin{split}
V_{11}&=\cos(wt/2\hbar)~,~~~~~~~~~V_{21}=\frac{s+\mu}{w}\sin(wt/2\hbar)\\
V_{31}&=i\frac{s-\mu}{w}\sin(wt/2\hbar)~,~V_{41}=\frac{-2ir\cos\theta}{w}\sin(wt/2\hbar)
\end{split}
\end{equation}

It does not matter what the other matrix elements in operator $V$ are, while we can determine the operator by Schmidt Orthogonalization under the constrain that the operator must be unitary. We can also decompose it into single and two-qubit operators just as shown in the dotted box of Fig. \ref{fig11:cir3.pdf}. Suppose that the normalized first column of the two-qubit operator is $\scriptsize [V_{11}^{'},V_{21}^{'},V_{31}^{'},V_{41}^{'}]$ satisfying normalization conditions $\scriptsize \sum_{i=1}^{4}\vert V_{i1}^{'} \vert ^{2}=1$,  and then we can construct 4$\times$4 unitary matrix $V$ that satisfies the constrain and the concrete form of operators can be determined by Eq.(\ref{qqq1}).
\begin{equation}
V=(V_{1}\oplus V_{2})\cdot (V_{0}\otimes I)
\label{qqq1}
\end{equation}
where the single-qubit unitary operators $V_{0}$ and two-qubit controlled operators $V_{k} (k=1,2)$ are determined by
\begin{equation}
\begin{split}
V_{0}&=\begin{pmatrix}
\sqrt{\vert V_{11}^{'}\vert ^{2}+\vert V_{21}^{'}\vert ^{2}}&\sqrt{\vert V_{31}^{'} \vert ^{2}+\vert V_{41}^{'}\vert ^{2}}\\
\sqrt{\vert V_{31}^{'}\vert ^{2}+\vert V_{41}^{'}\vert ^{2}}&-\sqrt{\vert V_{11}^{'}\vert ^{2}+\vert V_{21}^{'}\vert ^{2}}\\
\end{pmatrix}
=R(\theta_{0})
\\
V_{k}&=\begin{pmatrix}
\frac{V_{2k-1,1}^{'}}{\sqrt{\vert V_{2k-1,1}^{'}\vert ^{2}+\vert V_{2k,1}^{'}\vert ^{2}}}&\frac{V_{2k,1}^{'}}{\sqrt{\vert V_{2k-1,1}^{'}\vert ^{2}+\vert V_{2k,1}^{'}\vert ^{2}}}\\
\frac{V_{2k,1}^{'}}{\sqrt{\vert V_{2k-1,1}^{'}\vert ^{2}+\vert V_{2k,1}^{'}\vert ^{2}}}&\frac{-V_{2k-1,1}^{'}}{\sqrt{\vert V_{2k-1,1}^{'}\vert ^{2}+\vert V_{2k,1}^{'}\vert ^{2}}}\\
\end{pmatrix}
=R(\theta_{k})
\end{split}
\end{equation}

The rotation operator can be expressed as $\scriptsize R(\theta_{l})=\begin{pmatrix}\cos\theta_{l}&\sin\theta_{l}\\\sin\theta_{l}&-\cos\theta_{l}\end{pmatrix}$, $(l=0,1,2)$. According to the decomposition method and parameters in the generalized anti-$\mathcal{PT}$-symmetric Hamiltonian, we can determine the explicit forms of the angles in the three operators.

\begin{equation}
\begin{split}
\theta_{0}&=\arccos\sqrt{\frac{w^{2}\cos^{2}(wt/2\hbar)+(\mu+s)^{2}\sin^{2}(wt/2\hbar)}{w^{2}+2(\mu+s)^{2}\sin^{2}(wt/2\hbar)}}\\
\theta_{1}&=\arccos\frac{w\cos(wt/2\hbar)}{\sqrt{w^{2}\cos^{2}(wt/2\hbar)+(\mu+s)^{2}\sin^{2}(wt/2\hbar)}}\\
\theta_{2}&=\arccos\frac{i(s-\mu)}{\sqrt{(\mu-s)^{2}+4r^{2}\cos^{2}\theta}} \\
\end{split}
\end{equation}

The single-qubit operator $V_{0}$ and two controlled-$V_{k} (k=1,2)$ gates are all unitary, which is feasible to realize in quantum computation frame. Next, the three-qubit controlled-$U_{i} (i=1,2,3,4)$ construct a set of complete basis in two-dimensional Hilbert space on the work system and the construction method is not unique, which means operators $U_{i}$ can be simply set as identity matrix $\sigma_{0}=I_{2\times2}$ and Pauli matrix $\sigma_{x},\sigma_{y},\sigma_{z}$. Finally, two Hardmard gates are applied on the ancillary qubits to mix up the states and the target quantum state $\hat{\rho}(t)$ of the work system can be obtained by measuring the work system in the subspace where the ancillary qubits are $\ket{00}\bra{00}$ according to our parameters setup. It is worth emphasizing that our scheme works for both unbroken and broken anti-$\mathcal{PT}$-symmetric phase, even at the exceptional point. Therefore, our protocol provides a novel method to investigate various properties of anti-$\mathcal{PT}$-symmetric system and we apply it to experimental observations of information flow.  

\section{Experimental Implementation and Results}
To present the information retrieval in anti-$\mathcal{PT}$-symmetric system, we identify the information flow by trace distance between two quantum states 
\begin{equation}
\begin{split}
D(\hat{\rho}_{1}(t),\hat{\rho}_{2}(t))=\frac{1}{2}\textup{tr}\vert \hat{\rho}_{1}(t)-\hat{\rho}_{2}(t) \vert
\end{split}
\end{equation}
and $\vert \hat{M} \vert=\sqrt{\hat{M}^{\dagger}\hat{M}}$ \cite{flowpra,flowpra2}. The trace distance keeps invariant under unitary transformations whereas does not increase under completely positive and trace-preserving maps, which means the unidirectional information flow from the system to environment will not be recovered. However, the complete information retrieval from the environment in $\mathcal{PT}$-symmetric system has been proposed in theory \cite{flow}. In contrast, there has been little research in the counterpart anti-$\mathcal{PT}$-symmetric system and in this work, we observed an oscillation of information flow in anti-$\mathcal{PT}$-symmetric single-qubit system in experiment based on our Nuclear Magnetic Resonance platform. 

As a proof-of-principle experiment, we consider a two-level anti-$\mathcal{PT}$-symmetric system 
\begin{equation}
\begin{split}
H_{APT}=s(\textup{i}\hat{\sigma}_{x}+\lambda\hat{\sigma}_{z})
\label{expH}
\end{split}
\end{equation}
where $s\geq0$ is an energy scale and $\lambda \geq 0$ is a coefficient representing the degree of Hermiticity. We take different $\lambda$ values in both anti-$\mathcal{PT}$-symmetric unbroken and broken region to observe the dynamic feature of the system. According to the definition of information flow, we need to evolve the system under anti-$\mathcal{PT}$-symmetric dynamics with different initial state $\ket{0}\bra{0}$ and $\ket{1}\bra{1}$ to determine the distinguishability. Hence, we add one more working system in experiment ensuring that the two working qubits undergo the same anti-$\mathcal{PT}$-symmetric evolution just with different initial states. We use the spatial averaging technique to prepare the pseudo-pure state \cite{pps_space1,pps_space2} from the thermal equilibrium as the initial state and the form can be expressed as

\begin{equation}
\begin{split}
\ket{\rho_{0000}} = (1-\epsilon)I_{16}/2^{4}+\epsilon\ket{0000}\bra{0000}
\label{pps_state}
\end{split}
\end{equation}
where $I_{16}$ is a 16$\times$16 identity operator and $\epsilon\approx10^{-5}$ is the polarization. The first item does not evolve under unitary operators and the second deviated part is equivalent to quantum pure state. The fidelities between experimental and ideal pure state $\ket{0000}$ are over 99.5\%, which are calculated by the formula \cite{fidelity} 
\begin{equation}
\begin{split}
F(\rho,\sigma)=\text{tr}(\rho\sigma)/\sqrt{\text{tr}\rho^2}\sqrt{\text{tr}\sigma^2}
\end{split}
\end{equation}

\begin{figure}
         \centering
  	 \includegraphics[scale=0.33]{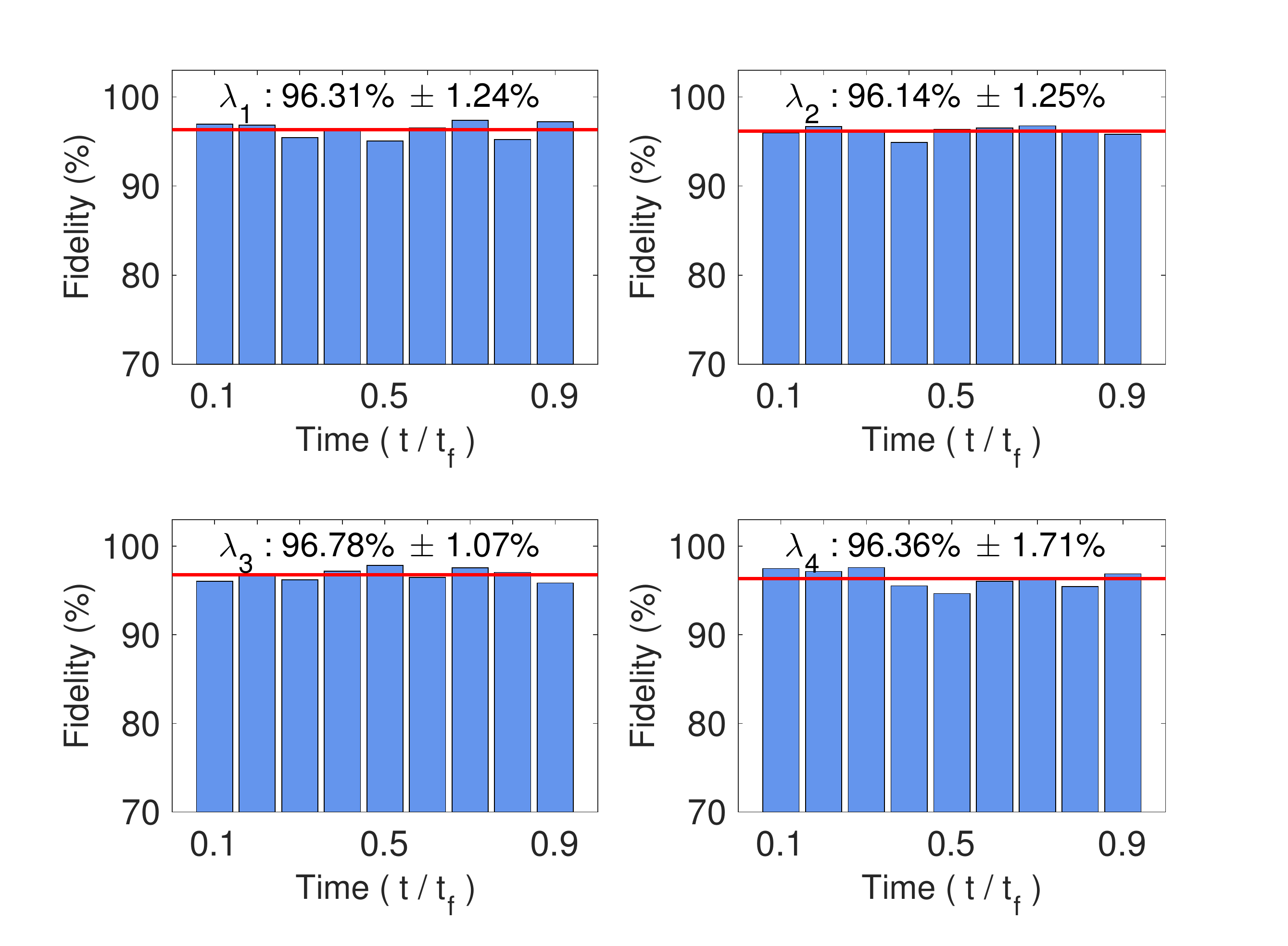}
          \caption{Fidelities between the experimental results and theoretical expectation at nine discrete time points of different $\lambda$ values. The average fidelities labelled by red lines are over 96\% and the maximum deviations are within 2\%.}
          \label{fig22:fid} 
\end{figure}

\begin{figure*}
 	\centering
  	  \includegraphics[scale=0.45]{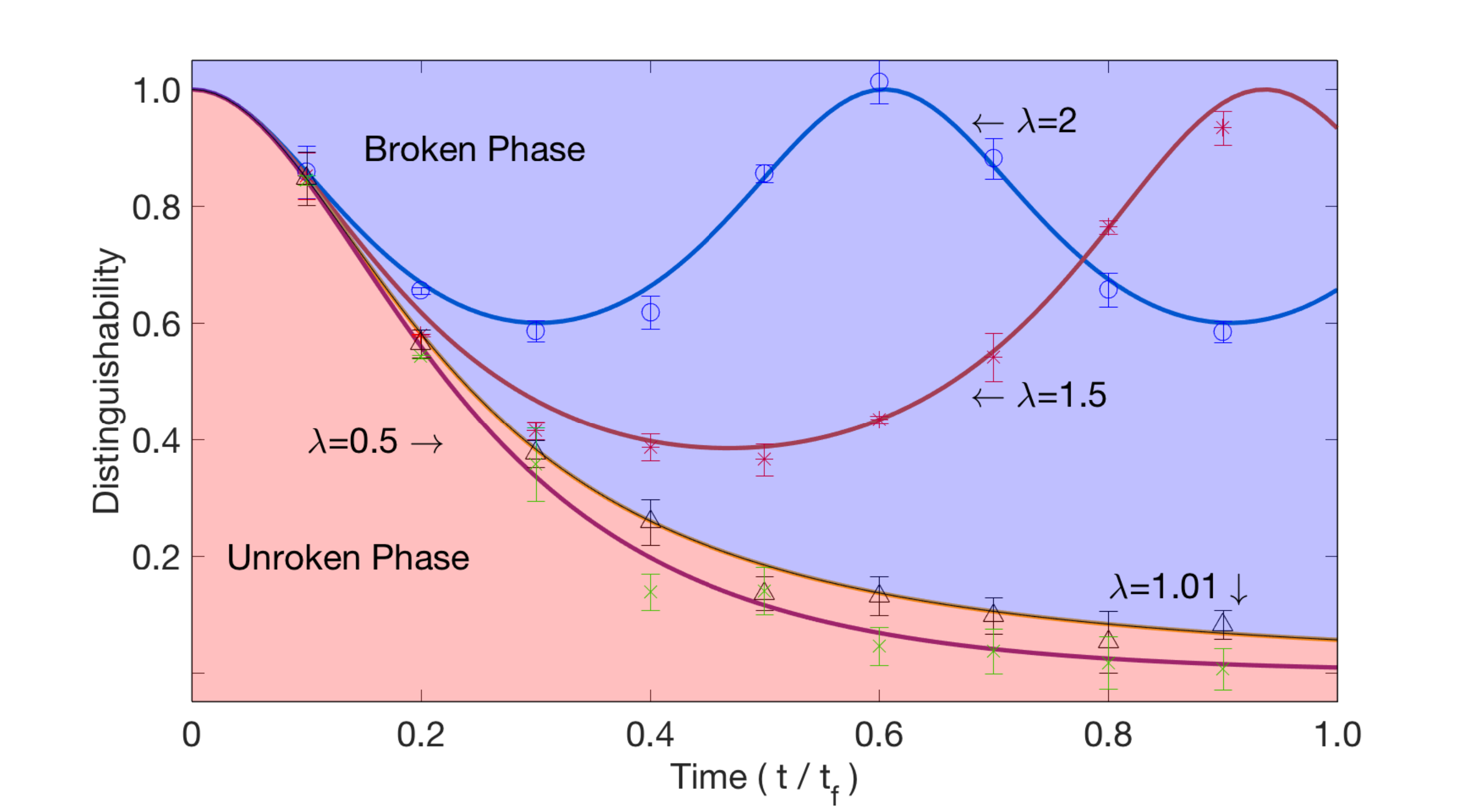}
          \caption{Experimental results of information flow measured by distinguishability. Four $\lambda$ values are set in our experiment including three broken anti-$\mathcal{PT}$-symmetric phase and one unbroken point. The solid lines represent theoretical values, and the nine discrete points on each line are experimental results. Theoretical error range of distinguishability are numerically analyzed based on the assumption that the fluctuations of amplitude and phase are within a range of 5\%.}
          \label{fig33:exp} 
\end{figure*}

Subsequently, we apply the quantum operations in our algorithm according to the different parameter setup. Four values of $\lambda \in \{2,1.5,1.01,0.5\}$ are chosen and the former three are located at broken phase and the last one leads to unbroken anti-$\mathcal{PT}$-symmetry. All the operations are realized using shaped pulses optimized by the gradient method \cite{grape1}. Each shaped pulse is simulated to be over 99.5\% fidelity while being robust to the static field distributions and inhomogeneity. By performing four-qubit quantum state tomography \cite{tomo1,tomo2,tomo3}, we obtain the target density matrix when the ancillary qubits are $\ket{00}$ at the end of circuit. We extract experimental data at nine discrete time points and the mean fidelities between the theoretical expectations and experimental values are over 96\% in Fig. \ref{fig22:fid}. The information flow identified by experimental results are plotted in Fig. \ref{fig33:exp}. Because of the random fluctuations of the amplitude and phase in control field, the experimental results produce some random errors.  We suppose that the random fluctuations are within a range of 5\% in amplitude and in phase, which are common in actual experimental process, then the fluctuation range of distinguishability are also plotted as the errorbar.

\begin{figure*}
          \centering
          \subfigure[]{
           \label{fig44:subfig:a}
  	  \includegraphics[scale=0.4]{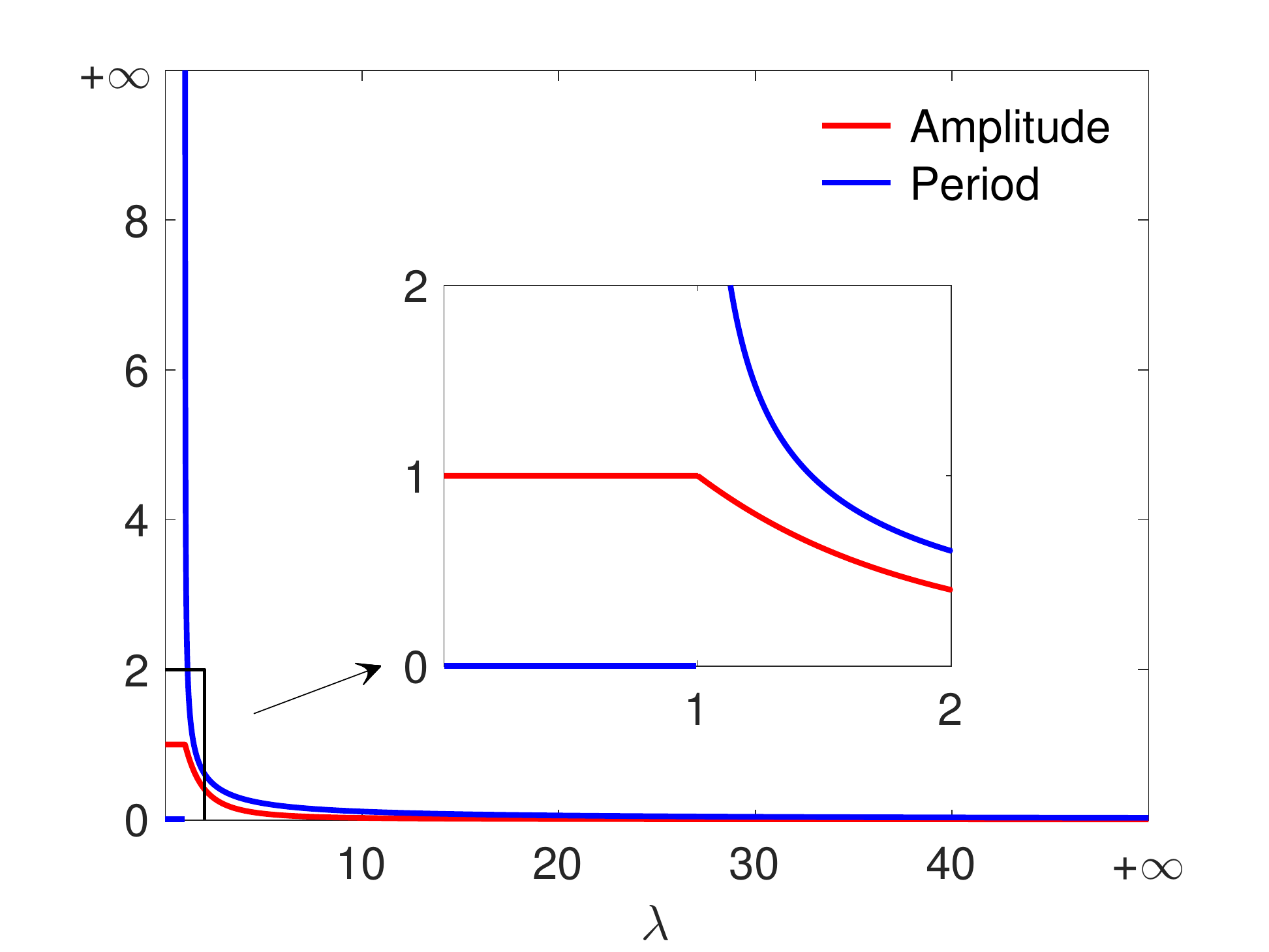}}
	    \subfigure[]{
           \label{fig44:subfig:b}
  	  \includegraphics[scale=0.4]{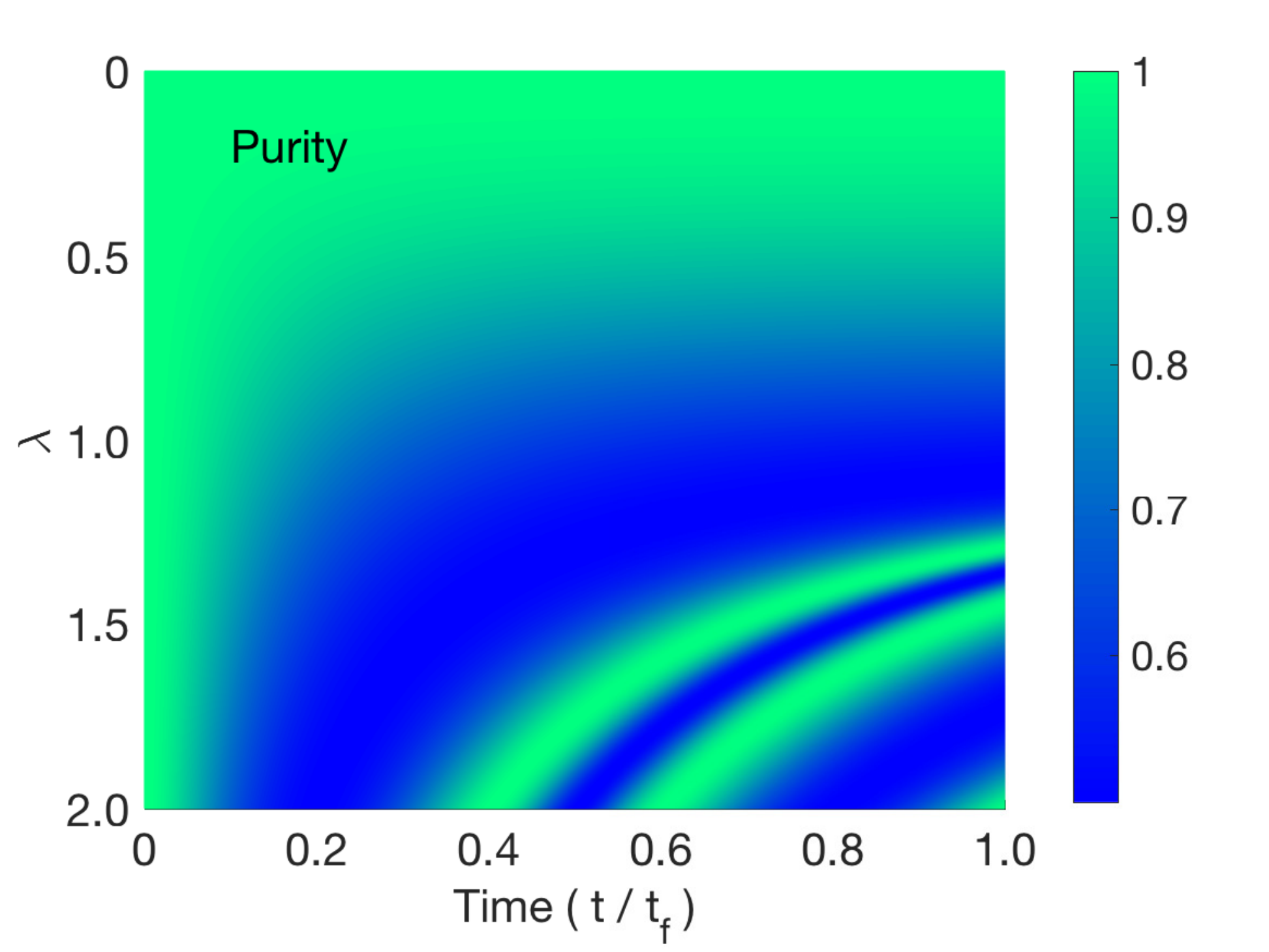}}
          \caption{(a) The amplitude and period of distinguishability as functions of $\lambda$ according the same parameter setup in experiment. The locally enlarged subgraph is the change trend around exceptional point. (b) Purity of quantum state evolving with time under different $\lambda$ values. When parameters locate in the unbroken anti-$\mathcal{PT}$-symmetric phase ($\lambda<1$), purity will decay exponentially with time and approach value 0.5, which means a maximum mixed state. The oscillation of purity can be observed in the broken phase.}
          \label{fig4:theory}
\end{figure*}

Our experimental results clearly show that the distinguishability oscillates with evolution time when the system symmetry is broken and information can retrieve from the environment completely. The closer you get to the exceptional point, the bigger the period gets and the larger the amplitude of information vibration becomes, which means the system undergoes larger fluctuations. However, the distinguishability decays with time and no information recover from the environment in the unbroken anti-$\mathcal{PT}$-symmetric phase. The distinguishability oscillates with period $T=\pi\hbar/(s\sqrt{\lambda^{2}-1})$ and in order to validate our experimental results about change trend, we theoretically analyze the oscillation period and amplitude just as shown in Fig. \ref{fig44:subfig:a}. Such a change trend means that if $\lambda$ is large enough, the distinguishability will maintain unchanged at value one, which corresponds to the physical explanation that the non-Hermitian part ($\textup{i}\hat{\sigma}_{x}$) in system Hamiltonian can be ignored compared with the Hermitian part ($\lambda\hat{\sigma}_{z}$). Then the evolution process of the system can be approximated to unitary evolution, in which the information flow does not oscillate. In the symmetry-unbroken phase, the amplitude keeps value one and the period is zero, which means the system will lose all the information into the environment and information backflow does not occur. Therefore, the behavior of the system can be consistently understood and interpreted, whether it is an Hermitian or an anti-$\mathcal{PT}$-symmetric system. In addition, the increase of distinguishability in the broken phase implies that the anti-$\mathcal{PT}$-symmetric system exhibits unique non-Markovian behavior as well \cite{flow,flowpra}. To further determine the evolution characteristics, we theoretically calculated the purity of the quantum state in Fig. \ref{fig44:subfig:b} and observed similar oscillation and attenuation phenomena just like information flow. This means that the quantum state evolving under the anti-$\mathcal{PT}$-symmetric Hamiltonian will be entangled with the environment so as to a decoherence process, but the process is reversible in the case of broken phase and irreversible in the unbroken phase.

\section{Conclusion}
We propose an algorithm for the simulation of generalized anti-$\mathcal{PT}$-symmetric system and observe an oscillation of information flow in the experiment. We compare the performance when system parameters approach exceptional point and change from anti-$\mathcal{PT}$-symmetric broken phase into unbroken phase. It is found that both the oscillation period and amplitude increase monotonically before the phase transition, whereas no information will be retrieved from the environment any more after passing the critical point, which means that we have also realized a symmetry breaking process in anti-$\mathcal{PT}$-symmetric system. The monotone correspondence relation in symmetry-broken phase implies that our results could supply a metric method to measure the degree of non-hermiticity ($\propto$ amplitude $\in [0,1]$) for a quantum system. The change tendency of the information flow obtained in our experiment could supply a consistent understanding method for the Hermitian and anti-$\mathcal{PT}$-symmetric systems. Our proposed scheme can be extended to high-dimensional cases and other quantum computing platforms. 

\section{Acknowledgements} 
This work was supported by the National Basic Research Program of China under Grant Nos.2017YFA0303700 and 2015CB921001, National Natural Science Foundation of China under Grant Nos.61726801, 11474168 and 11474181. C. Z is supported by National Natural Science Foundation of China Grant No. 11705004, Open Research Fund Program of the State Key Laboratory of Low-Dimensional Quantum Physics No. KF201710. T. X is also supported by the Science, Technology and Innovation Commission of Shenzhen Municipality (No. ZDSYS20170303165926217, No. JCYJ20170412152620376) and Guangdong Innovative and Entrepreneurial Research Team Program (Grant No. 2016ZT06D348). We would thank Huawei Technology for financial support.

\appendix
\section{Experimental Setup}
\begin{figure*}
          \centering
 	 \subfigure[]{
 	  \label{fig55:subfig:a}
  	  \includegraphics[scale=0.4]{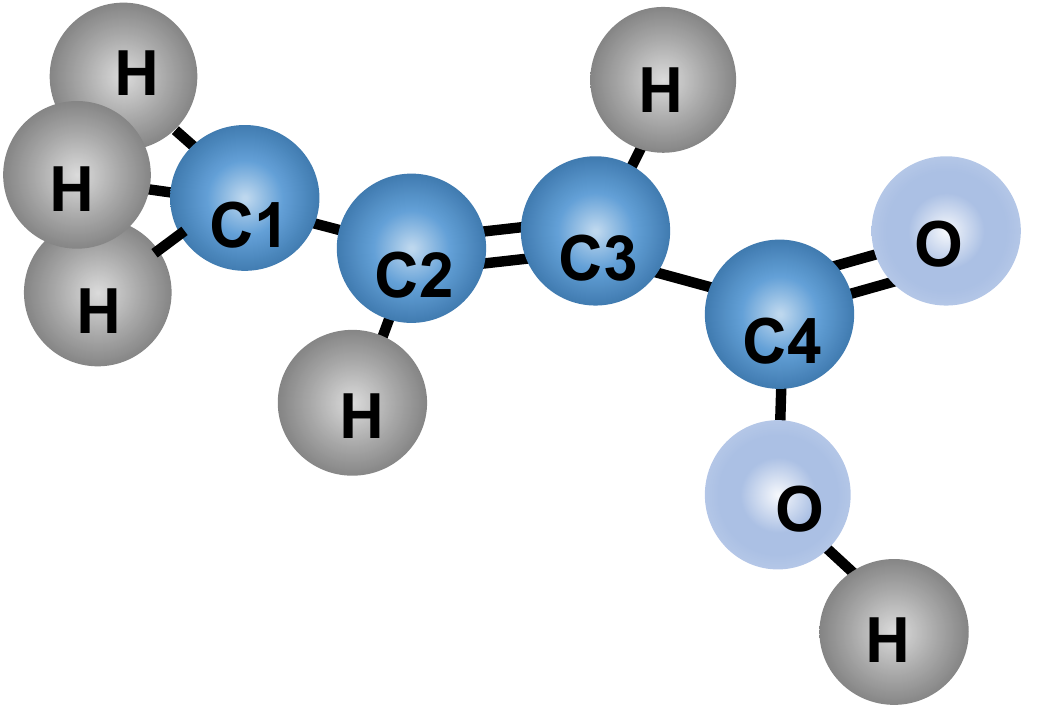}}
 	  \subfigure[]{
          \label{fig55:subfig:b} 
  	  \includegraphics[scale=0.3]{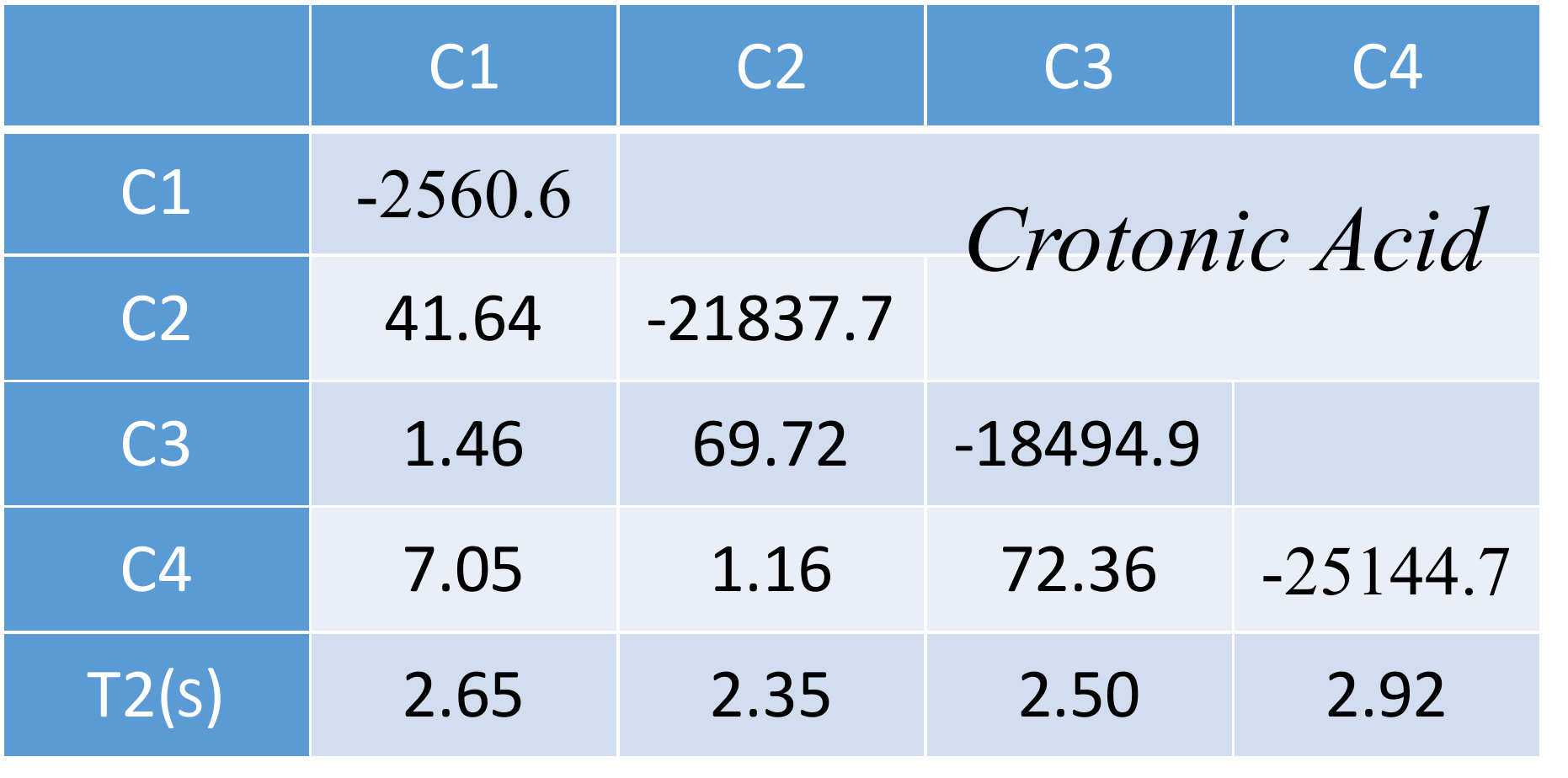}}
 	  \subfigure[]{
          \label{fig55:subfig:c} 
  	  \includegraphics[scale=1]{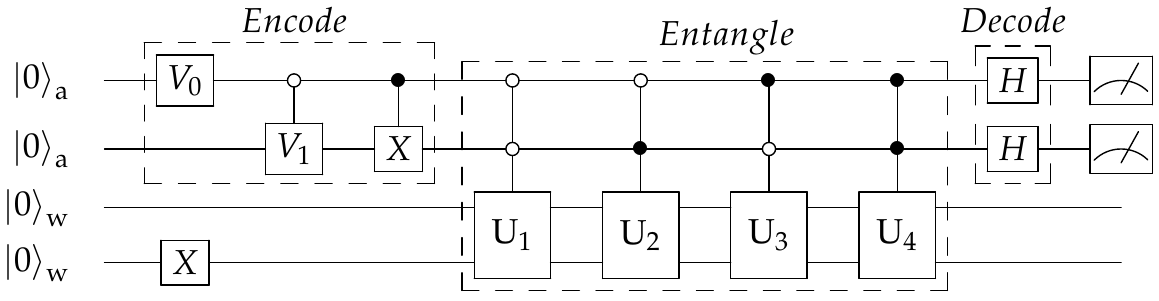}}
          \caption{(a) Molecule structure of ${}^{13}$C-labeled crotonic-acid. C1, C2, C3 and C4 are used as four qubits in the experiment, while all ${}^{1}H$ are decoupled throughout the experiment. (b) Molecule parameters of sample: the chemical shifts and J-couplings (in Hz) are listed by the diagonal and off-diagonal elements, respectively. Transversal relaxation time T2 (in seconds) are also shown at bottom. (c) Quantum circuit used to observe information flow in experiment. The first two qubits are used as ancillary qubits $\ket{00}_{\textup{a}}$ and the other qubits are working system $\ket{00}_{\textup{w}}$. After a series of unitary quantum gates, we measure the state of the working qubits when the ancillary qubits are $\ket{00}\bra{00}$.}
          \label{fig55:molecule}
\end{figure*}

In experiments, we use ${}^{13}C$-labeled transcrotonic acid dissolved in d6-acetone as the four-qubit sample. The structure of this molecule is shown in Fig. \ref{fig55:subfig:a}. Notations C1 to C4 denote the four qubits that we can control and C1, C2 are chosen as the ancillary qubits, C3 and C4 as the system qubits, respectively. All the ${}^{1}H$ are decoupled throughout the experiment. The experiments were carried out on a Bruker ADVANCE 600 MHz spectrometer at room temperature (298 K). The internal Hamiltonian of the sample under weak coupling approximation is 
\begin{equation}
\begin{split}
H_{int}=-\sum^{4}_{i=1}\pi\nu_{i}\sigma^{i}_{z}+\sum^{4}_{i<j}\frac{\pi}{2}J_{ij}\sigma^{i}_{z}\sigma^{j}_{z}
\end{split}
\end{equation}
where $\nu_{i}$ is the chemical shift and $J_{ij}$ is the J-coupling strength between the $i$th and $j$th nuclei. In Fig. \ref{fig55:subfig:b}, the molecular parameters are listed in the diagonal and off-diagonal elements. 
 
\textbf{Initialization:}
The process of quantum computation in liquid Nuclear Magnetic Resonance system starts from a thermal equilibrium state obeying Boltzmann distribution at room temperature $T$:
\begin{equation}
\begin{split}
\rho_{eq}=\frac{e^{-H_{int}/k_{B}T}}{Tr(e^{-H_{int}/k_{B}T})}
\end{split}
\end{equation}
where $k_{B}$ is the Boltzmann constant \cite{NMRQ}. It's normal that $\| H_{int}/k_{B}T \|$ is much less than one and $J_{kl}\ll \omega_{i}$, so the thermal equilibrium state can be approximated as 

\begin{equation}
\rho_{eq} \approx \frac{1}{2^{4}}(I^{\otimes 4}+\sum_{i}^{4}\frac{\hbar w_{i} \sigma_{z}^{i}}{2k_{B}T}) \label{eq2}
\end{equation} 
where the notation $I$ is identity matrix and $\sigma_{z}$ is a Pauli matrix. To initialize the system, we generally need to drive the quantum system from the highly mixed state $\rho_{eq}$, which can not be used as an initial state to the pseudo-pure state in Eq.(\ref{pps_state}). This is realized via spatial averaging technique \cite{pps_space1,pps_space2}, where all the processes are realized by unitary operations including single-qubit rotations, controlled-NOT gates and gradient fields in the $z$ direction. Full quantum state tomography \cite{tomo1,tomo2,tomo3} is then performed in order to obtain a quantitative estimation of the quality of our pseudo-pure state. We found that the fidelity between the prepared pseudo-pure state and the target state is over 99\% and this state serves as the starting point for subsequent computation tasks.

\begin{figure*}
          \centering
 	 \subfigure{
 	  \label{fig66:subfig:a}
  	  \includegraphics[scale=0.4]{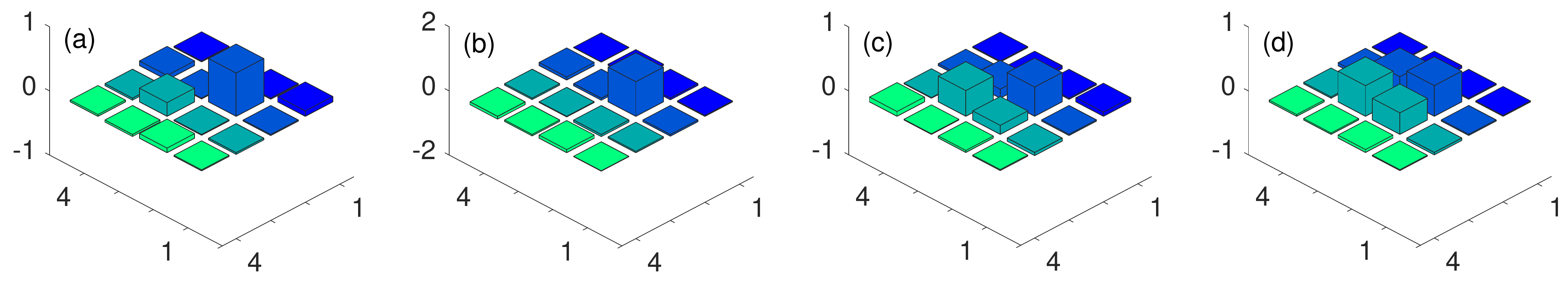}}
	  \subfigure{
 	  \label{fig66:subfig:b}
  	  \includegraphics[scale=0.4]{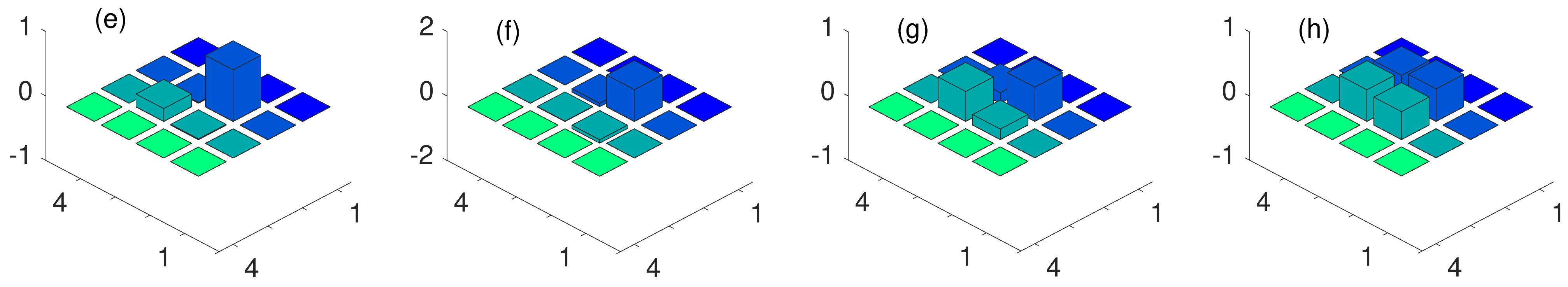}}
          \caption{Real parts of quantum state of work system at the ninth time points. Figures in the first row from (a) to (d) represent the real part of the final experimental density matrix ($\lambda_{1} \to \lambda_{4}$) and the corresponding theoretical values are shown on the following row from (e) to (h).}
          \label{fig66:rho}
\end{figure*}

\textbf{Readout:}
 The measurement in Nuclear Magnetic Resonance detection is performed on a bulk ensemble of molecules, which means the readout is an ensemble-averaged macroscopic measurement. At the end of the quantum circuit, all experimental data are extracted from the free-induction decay (FID), which is the signal induced by the precessing magnetization of the sample in a surrounding detection coil. FID is recorded as a time-domain signal, which consists of a number of oscillating waves of different frequencies, amplitudes, and phases. The signal is then subjected to Fourier transformation, and the resulting spectral lines are fitted, yielding a set of measurement data \cite{NMRQ}. We obtain the final density matrix by performing quantum state tomography. It is finished by applying 17 readout pulses with a duration of 0.9 ms after the evolution. Then we can reconstruct all the density matrix elements of the final state. We find the subspace where the ancillary qubits are state $\ket{00}\bra{00}$ to get the target quantum state of the work system. The real parts of the density matrices of work system for the experimental results at the last time point under different $\lambda$ values and the corresponding theoretical values are displayed in Fig. \ref{fig66:rho} to evaluate the performance of our experiment.

\section{Experimental Protocol}
In experiment, the eigenvalues of the anti-$\mathcal{PT}$-symmetric Hamiltonian $H_{APT}$ in Eq.(\ref{expH}) are $\pm s\sqrt{\lambda^{2}-1}$ and the anti-$\mathcal{PT}$-symmetry will be unbroken if $\lambda<1$, and the two-order exceptional point is located at $\lambda=1$. The dynamic evolution under $H_{APT}$ can be realized via our protocol presented above by setting the parameters in Eq.(\ref{eq1}) as $\theta=0$ and $r=\lambda s=\lambda \mu$ appropriately. We fix evolution time $t_{f}=1s$ and set $s=3$ as the energy scale. To reduce the random error caused by the change of environment in experiment, we extend the controlled gates $U_{i}=\sigma_{i}$ to two-qubit Hilbert space $U_{i}=\sigma_{i}\otimes\sigma_{i}$ ($i \in \{0,x,y,z\}$) and add one more rotation $\sigma_{x}$ on the second work qubit. Therefore, the quantum systems in two different Hilbert spaces undergo the same dynamic evolution induced by the anti-$\mathcal{PT}$-symmetric Hamiltonian only with different initial state by this experimental setup.

Qauntum circuit that is used to observe information flow in the experiment is shown in the Fig. \ref{fig55:subfig:c}. Single-qubit opeartor $V_{0}$ and two-qubit opeartor $V_{1}$ are parameter-dependent quantum gates, while the other unitary quantum gates includig the controlled-$U_{i}$ on the working system don not vary with the parameters in the anti-$\mathcal{PT}$-symmetric Hamiltonian. Quantum evolution according to the quantum circuit we constructed are optimized by gradient ascent pulse engineering (GRAPE) \cite{grape1} with fidelity over 99.5\% and the durations of the optimized pulses are within 60 ms in experiment.

\bibliographystyle{unsrtnat}
\bibliography{antiPT_wen.bib}

\end{document}